\documentclass[preprint,3p, onecolumn]{elsarticle}

\usepackage{lineno}
\modulolinenumbers[1]

\usepackage[utf8x]{inputenc}
\usepackage{textgreek}
\usepackage{isotope}
\usepackage{subcaption}
\usepackage{mathtools} 
\usepackage{braket}
\usepackage{comment}
\usepackage[colorlinks,citecolor=blue,urlcolor=blue,linkcolor=blue]{hyperref}

\makeatletter
\def\ps@pprintTitle{%
	\let\@oddhead\@empty  
	\let\@evenhead\@empty
	\def\@oddfoot{\reset@font\hfil\thepage\hfil}
	\let\@evenfoot\@oddfoot
}
\makeatother

\begin{document}
\begin{frontmatter}

\title{On the multipole mixing ratio of the  $1066$ keV transition \\ from the 0.52 $\mu s$ isomer of $^{180}$Hf}


\author[mymainaddress]{A.~Chalil\corref{mycorrespondingauthor}\footnote{Present Address: Universit\'e Claude Bernard Lyon 1, CNRS/IN2P3, IP2I Lyon, F-69622 Villeurbanne, France}}
\ead{a.chalil@ip2i.in2p3.fr}
\author[mymainaddress]{T.~J.~Mertzimekis}
\author[mymainaddress]{A.~Zyriliou}
\author[mymainaddress]{P.~Vasileiou}
\author[mysecondaryaddress]{N.~Florea}
\author[mysecondaryaddress]{C.~Mihai}
\author[mysecondaryaddress]{R.~Mihai}
\author[mysecondaryaddress]{C.~Nita}
\author[mysecondaryaddress]{C.~Sotty}
\author[mysecondaryaddress]{A.~Mitu}
\author[mysecondaryaddress]{L.~Stan}
\author[mysecondaryaddress]{A.~Turturica}
\author[mysecondaryaddress]{R.~M\u{a}rginean}
\author[mysecondaryaddress]{N.~M\u{a}rginean}


\cortext[mycorrespondingauthor]{Corresponding author}

\address[mymainaddress]{National and Kapodistrian University of Athens, Department of Physics, GR-15784, Athens, Greece}
\address[mysecondaryaddress]{National Institute of Physics and Nuclear Engineering, Magurele, Romania}

\begin{abstract}

The nucleus \isotope[180]{Hf} is one of the most primary of examples of an axially symmetric prolate rotor. Combined with the presence of high-$K$ isomers, spectroscopic studies can provide important information on the nature of its single-particle levels. Precise measurements are essential for constraining nuclear models and interpreting the nature of such isomeric states. In this work, the nucleus \isotope[180]{Hf} was populated using the proton pick-up reaction
\isotope[181][]{Ta}(\isotope[11]{B},\isotope[12]{C})\isotope[180]{Hf} at beam energy of 47 MeV at Horia Hulubei National Institute of Nuclear Physics and Engineering (IFIN-HH). The spin of the 1374 keV state and the mixing ratio of the $1066$ keV transition have been measured, the latter with an increased precision compared to the previous value from literature. The presently measured spin of the 1374 keV state, currently assigned a tentative value of $(4^-_1)$, favors one of the two different values reported in the literature. The particular state constitutes the band-head of a rotational band in \isotope[180][]{Hf}. The measured multipolarity mixing ratio of the inter-band transition $1374 \rightarrow 309$ keV can provide important information for the testing and constraining of theoretical nuclear models used for the study of the intrinsic properties of \isotope[180][]{Hf} as well as its neighboring isotopes.

\end{abstract}


\end{frontmatter}

\section{Introduction}
\label{intro}
The nucleus \isotope[180]{Hf} is located in a region where both protons and
neutrons typically occupy high--$\Omega$ orbitals located near the Fermi
surface~\cite{Tandel_2016_PhysRevC.94.064304}. This results in the presence
of high-$K$ isomers~\cite{Dracoulis_2016,Walker_2015}, which along with the associated rotational bands can provide important information about the underlying single-particle orbitals. The hafnium isotopes are some of the best
examples of rigid and axially symmetric prolate rotors, making the projection
$K$ a good quantum number in their description. 
\begin{figure}[ht]
	\centering
	\includegraphics[width=0.60\textwidth]{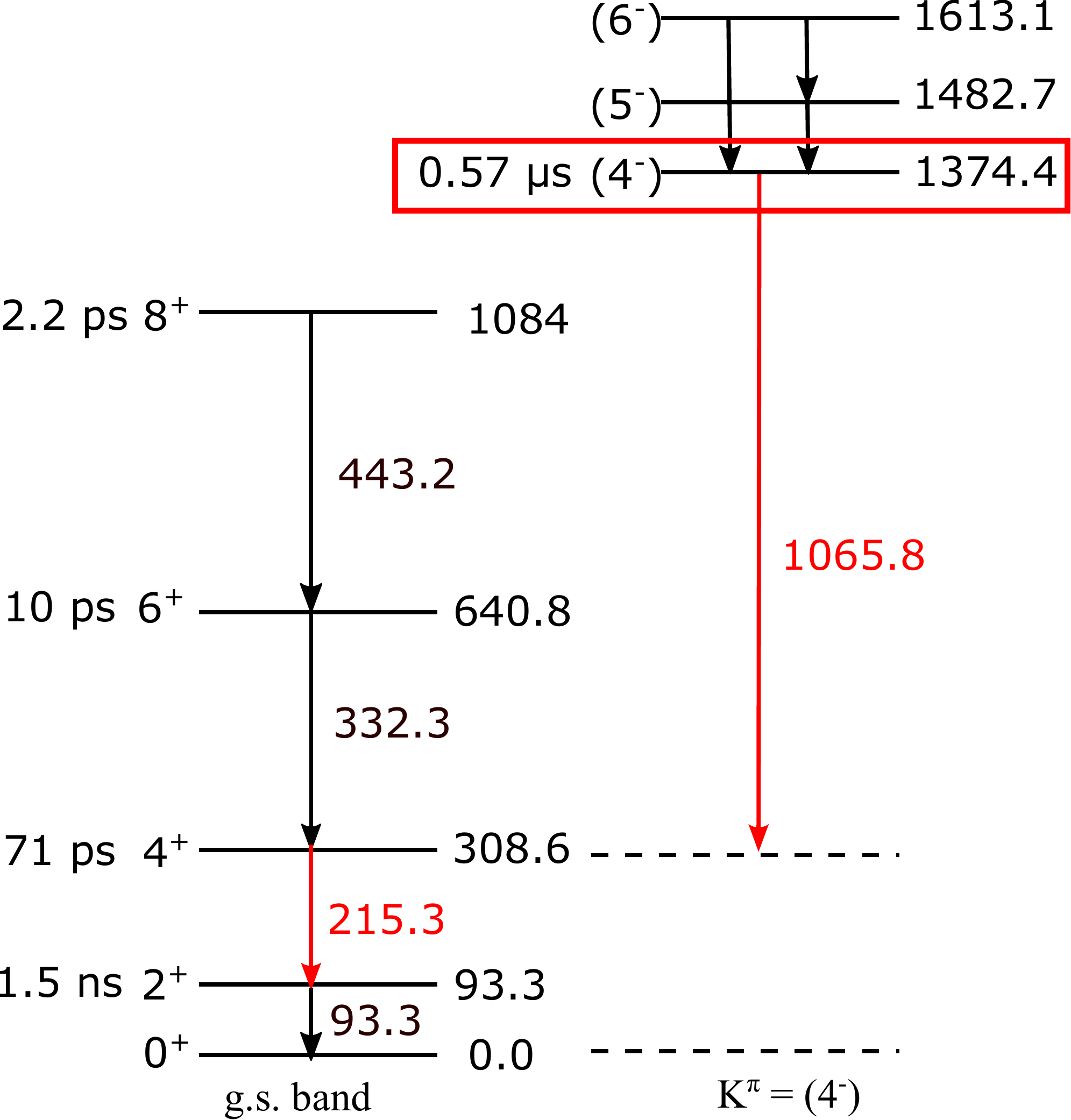}
	\caption{Partial level scheme of \isotope[180]{Hf}~\cite{NNDC}, showing the ground-state band and the $K^{\pi}=(4^-)$ bands. Red arrows indicate the measured two-$\gamma$ cascade in the present work. The mixing ratio of the transition $1374 \rightarrow 309$ keV is studied in this work with the angular correlation method. Energy units are in keV.}
	\label{fig:level_scheme_180Hf}
\end{figure}
The low-lying $K^\pi = 8^-$ isomer has been firstly studied 
in~\cite{Burson_1951_PhysRev.83.62} using activation experiments at Argonne
National Laboratory. In~\cite{Dalarcao_1999_PhysRevC.59.R1227}, pulsed
\isotope[238]{U} and \isotope[208]{Pb} beams have been used to populate
high-$K$ isomers in Hf isotopes, reporting $(10^+)$, $12^+$, $14^+$
and $(18^-)$ \textit{K}-isomers. In~\cite{Tandel_2016_PhysRevC.94.064304}, 
the number of \textit{K}-isomers is extended, reporting the $4^-,6^+$
\textit{K}-isomers, using centroid--shift and decay measurements. A recent
experimental campaign at IFIN-HH~\cite{Wiederhold_2019_PhysRevC.99.024316}
resulted in measurements of the lifetimes of the ground-state band up to
the $6^+_1$, employing the fast-timing technique. 

Still however, there is a significant amount of certain states and transitions with uncertain 
fundamental structure properties, either in terms of their spins, parities
and mixing ratios or their lifetimes and transition probabilities. Concerning spin assignments and multipolarity mixing ratios, of which the present work is focused, existing measurements on mixing ratios for the nucleus $^{180}$Hf feature quite large uncertainties, with very few exceptions, as seen in~\cite{NNDC}, while in a lot of cases this mixing remains unknown. For the case of the 1374 keV state (Fig.~\ref{fig:level_scheme_180Hf}), which is a $K$-isomer with a lifetime of $0.57~\mu s$~\cite{Tandel_2016_PhysRevC.94.064304,Grigorev1991}, is currently assigned a tentative spin value of $(4^-)$~\cite{NNDC}. Two measurements have been reported~\cite{Grigorev1991,Boneva1990} with different values for its spin. Data obtained from the GAMMASPHERE array~\cite{SIMON2000205} assume a definite spin-parity of $4^-$ for the particular state, although there is no DCO (Directional Correlations from Oriented states~\cite{KRANE_steffen_wheeler_1973351}) ratio in the table given in~\cite{Ngijoi_2007_PhysRevC.75.034305}, probably because of the unreliability of extracting DCO values from transitions which depopulate states with very long half-lives. The multipolarity mixing ratio of the depopulating inter-band transition $1374 \rightarrow 309$ keV measured in~\cite{Grigorev1991} features a quite large relative uncertainty of 250\%. There is thus sufficient motivation for new measurement, in an attempt to establish the spin value of the 1374 keV state and provide a more precise measurement of its depopulating transition to the $4^+_1$ state of the ground-state band. It has to be stated also that the 1374 keV state is a band-head of the currently assigned $K^{\pi}=(4^-)$ rotational band, thus its structure properties are affecting all levels that belong to this particular band.
 
The study of the de-exciting $\gamma$-rays of excited nuclei can offer valuable information about the structure of the depopulated states. Using the angular correlation method, the multipolarity mixing ratio, $\delta$, can be determined experimentally and thus establish the degree of mixing between the participating multipoles of the transition~\cite{Rose_Brink_Revmodphys_1967,Hamilton1975_electromagnetic,SMITH201947,Bernards_2011_PhysRevC.84.047304,Goodin_2008_PhysRevC.78.044331}.

The mixing ratio is defined as the ratio of the transition strength of the two lowest multipoles $L, L'$ allowed for $\gamma$-decay~\cite{KRANE1975_mixing_ratio_383}:
\begin{equation}\label{eq: mixing ratio}
		\delta= \frac{\braket{||L'||}}{\braket{||L||}} = \frac{\braket{||L+1||}}{\braket{||L||}}.
\end{equation}

Precise mixing ratio values are highly important, as they can determine
the degree of partitioning of the multipolarity of $\gamma$ radiation~\cite{KRANE1975_mixing_ratio_383}. This
level of partitioning is an important input for the calculation of reduced
transition probabilities of $\gamma$ transitions~\cite{Lange_1982_RevModPhys.54.119}. In the present work, the proton--pickup reaction
\begin{center}
\isotope[181]{Ta}(\isotope[11]{B},\isotope[12]{C})\isotope[180]{Hf}
\end{center}
has been employed to populate several states from the ground--state
and side bands. A more precise measurement of the mixing ratio of the transition $1374 \rightarrow 309$ keV, or in terms of their adopted spins values $(4^-_1) \rightarrow 4^+$, is presented in this work and compared with previous spectroscopic studies~\cite{Grigorev1991,Boneva1990}. The lifetime of the depopulated initial state is long enough $(\gg 1~ns)$ to allow for the loss of the initial spin alignment coming from the reaction~\cite{Matthias_1965_PhysRevLett.14.46}. Thus, the state can be safely considered randomly oriented and the usual formalism of $\gamma$-directional correlations can be applied~\cite{KRANE_steffen_wheeler_1973351,Rose_Brink_Revmodphys_1967,Hamilton1975_electromagnetic, Biedenharn_Rose_RevModPhys.25.729, STUCHBERY_2002753, Chalil2022}, as long as the intermediate state's lifetime is short enough, in order to measure the mixing ratio of the depopulating transition.

\section{Experimental details}
\label{experimental}
\begin{figure*}[t]
	\centering
	\includegraphics[width=0.88\textwidth]{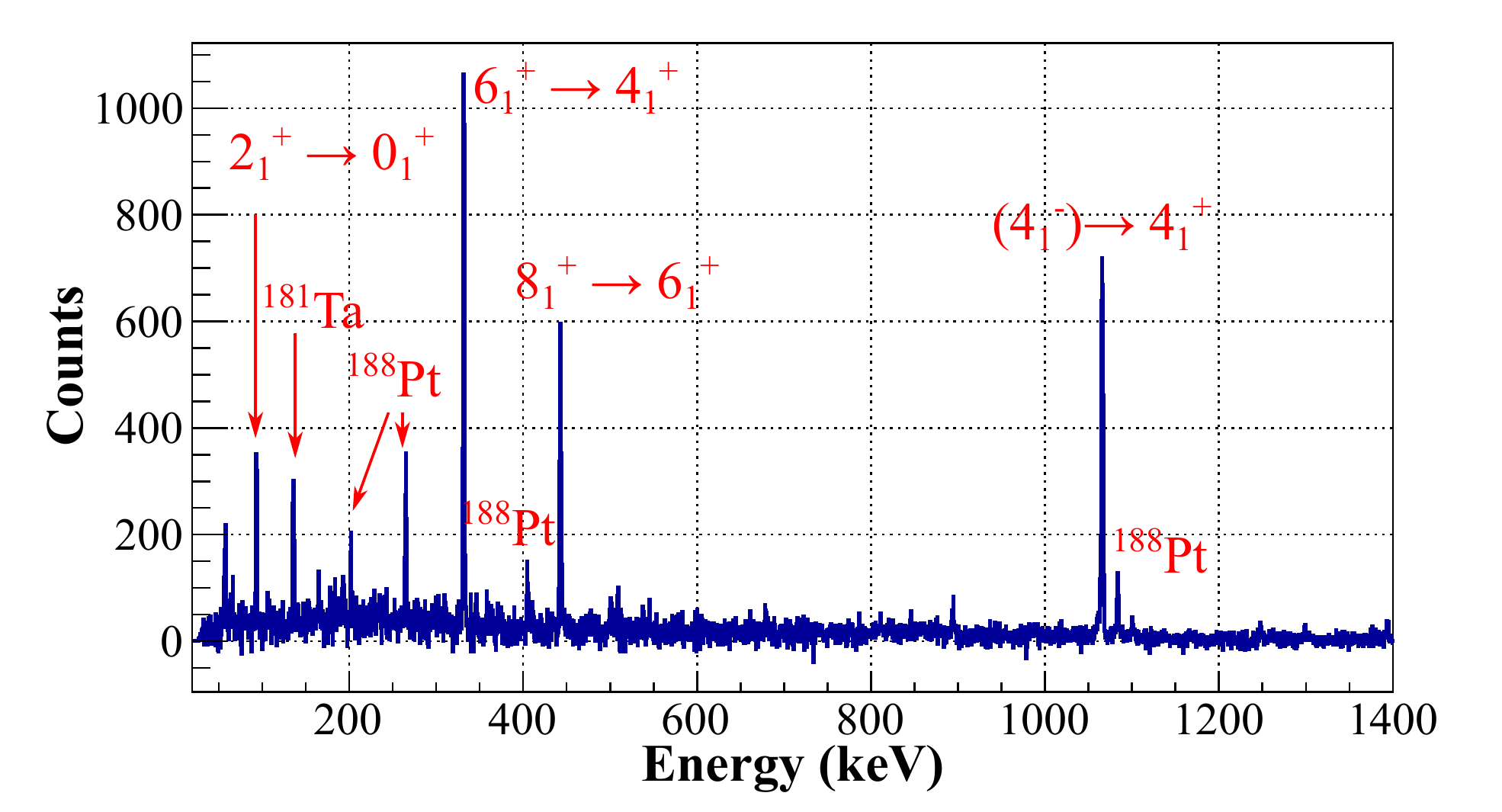}
	\caption{Projection of the $\gamma \gamma$ matrix gated at the 215 keV g.s. band transition. The relative angle between the detector pairs is $40$ deg. The ground-state band transitions up to the level $8^+_1$ are shown, as well as the studied inter-band transition $(4_1^-) \rightarrow 4^+_1$. Contaminant peaks coming from fusion-evaporation reactions (\isotope[188][]{Pt}) and Coulomb excitation on the target (\isotope[181][]{Ta}) are not overlapping with the photopeaks of interest.}
	\label{fig: spectrum}
\end{figure*}
The experiment was performed at the 9MV Tandem accelerator of the Horia Hulubei National Institute of Nuclear Physics and Engineering (IFIN-HH). The nucleus \isotope[180][]{Hf} was populated using the proton pick-up reaction \isotope[181][]{Ta}(\isotope[11][]{B},\isotope[12][]{C})\isotope[180][]{Hf} at a beam energy of $47$ MeV. A 5~mg$\cdot$cm$^{-2}$ metallic \isotope{Ta} target was used in this work, with a natural abundance of 99.99\% in \isotope[181]{Ta}~\cite{NNDC}. The $\gamma$ rays emitted during the decay of the reaction products were measured by the ROSPHERE
array~\cite{Bucurescu_2016_1}, consisting of 25 HPGe detectors for the reported measurement. A total of $7 \times 10^8$ events, with the trigger set to record an event when at least two HPGe detectors fire were collected during
the three-day run time of the experiment. The two-fold events were chosen for further
analysis, as they have exhibited reduced background from fusion-evaporation reactions.
These events were then sorted into three--dimensional cubes with the relative angle
as the index number, i.e. $\gamma-\gamma-\theta_{rel}$ cubes. All possible pairs of HPGe detectors with a specific relative
angle are then grouped together in 5 relative angles below $90^{\circ}$: $21^{\circ}, 40^{\circ}, 60^{\circ}, 72^{\circ}$ and $81^{\circ}$. In Fig.~\ref{fig: spectrum}, a projection of a $\gamma \gamma$ matrix is shown, after gating on the $4^+_1 \rightarrow 2^+_1$ ($E_\gamma = 215$ keV) transition, where the detector pairs' relative angle is $40$ deg. The transitions between the low-lying states are shown clearly, and contaminations from photopeaks coming isotopes produced from fusion-evaporation reaction (e.g. \isotope[188]{Pt}) do not overlap with the transitions of interest.

 The detector efficiencies have been measured for all 25 HPGe detectors of the
current setup of ROSPHERE. The efficiency correction has been determined using two
\isotope[152]{Eu} sources before and after the experiment and was implemented
after the sorting of the data, using the formula~\cite{Patel_2002}:
\begin{equation}
    \epsilon(E_1^\gamma,E_2^\gamma,\theta_{rel})= \frac{1}{2}\sum_{i \neq j} \left[ \epsilon_i(E_1^\gamma) \epsilon_j(E_2^\gamma) + \epsilon_i(E_2^\gamma) \epsilon_j(E_1^\gamma) \right],
\end{equation}
where $\epsilon_i,\epsilon_j$ are the efficiencies of the detectors with relative
angle $\theta_{rel}$ and $E_1^\gamma,E_2^\gamma$ are the energies of the $\gamma$
transitions involved in the cascade. The $1/2$ factor is used to prevent the double
counting between the same pairs of efficiencies.

\begin{figure*}[t]
\centering
	\begin{subfigure}{0.48\textwidth}
		\includegraphics[width=\textwidth]{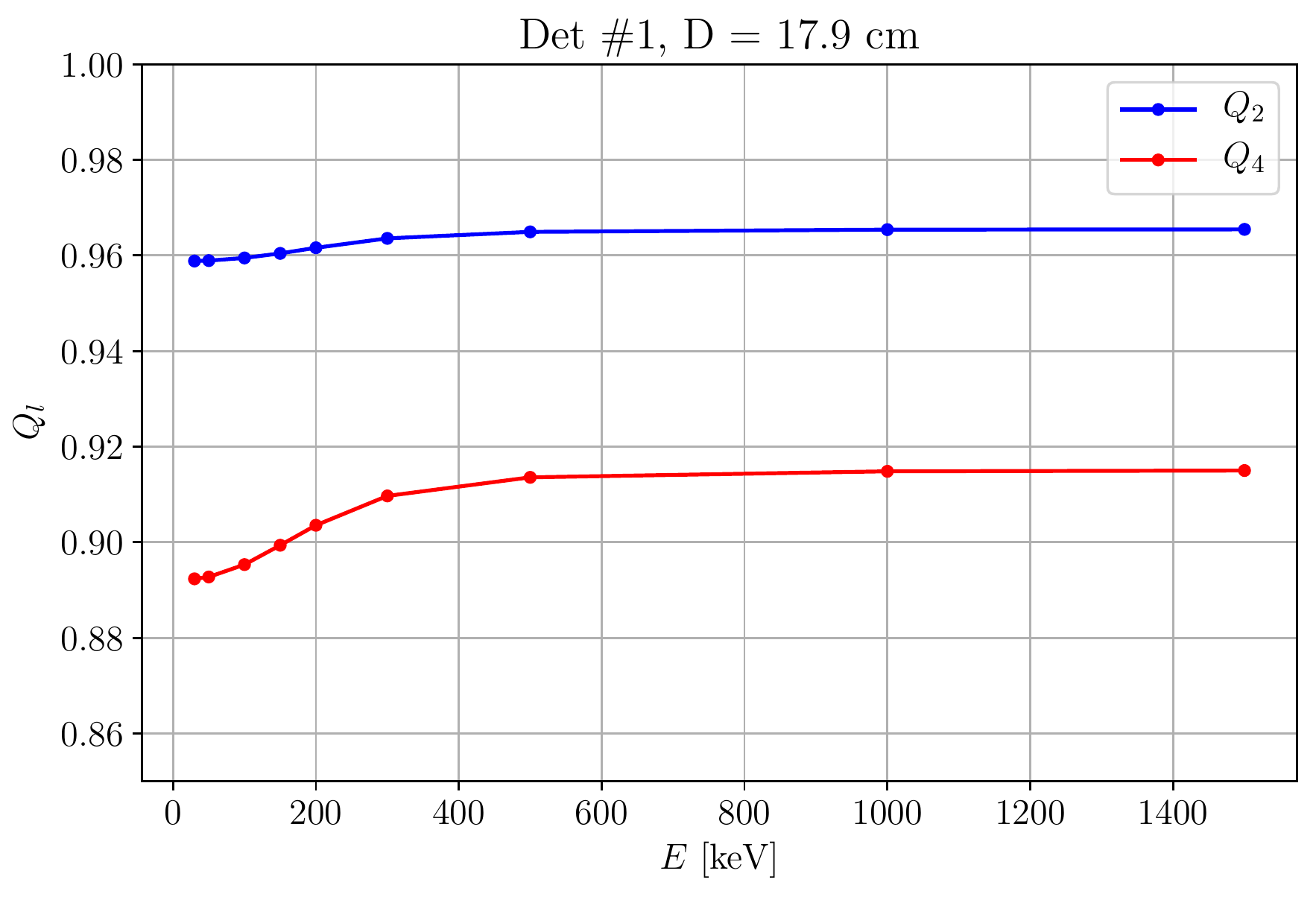}
		\centering
	\end{subfigure}
	\begin{subfigure}{0.48\textwidth}
		\includegraphics[width=\textwidth]{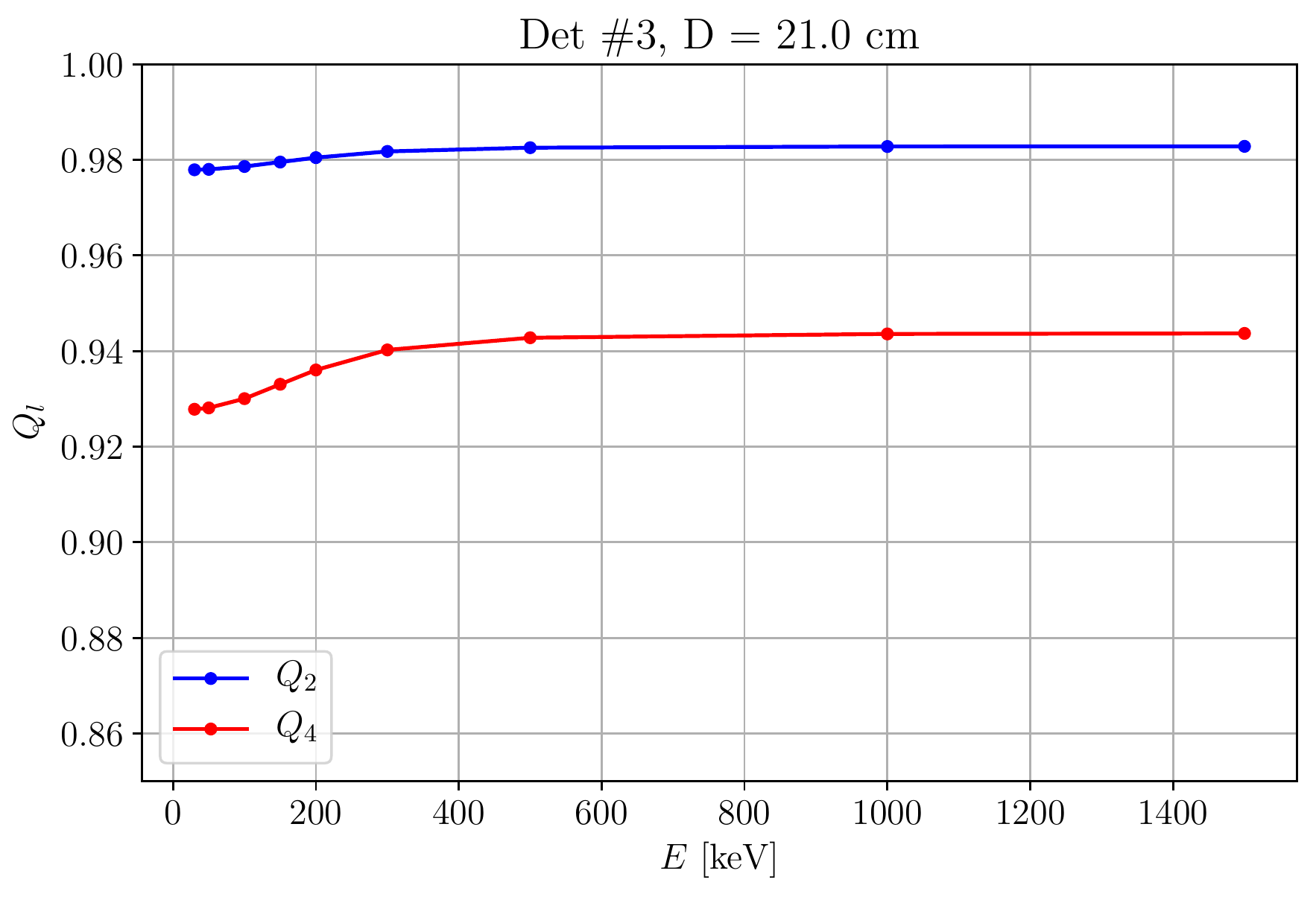}
		\centering
	\end{subfigure}%
	\caption{Typical behavior of the geometrical attenuation coefficients for two (2)
	out of twenty five (25) HPGe detectors of ROSPHERE as a function of $\gamma$ energy. The target-to-detector distance is shown on the top of the figures.}
	\label{fig:geometrical_corrections}
\end{figure*}

The finite size of the detectors can have an important effect on angular correlation measurements~\cite{Krane1973SolidangleCF,KRANE1972_solid_angle_205,BARRETTE19711_solid_angle}.
The effect of the detectors' dimensions should be incorporated in the angular correlation
function in order to determine spins and mixing ratios with significant
accuracy. A Python~\cite{python36} program has been developed based on Krane's
formalism~\cite{Krane_1972_205} in order to incorporate appropriate corrections for
the ROSPHERE detectors~\cite{Bucurescu_2016_1} featuring coaxial geometry. In
Fig.~\ref{fig:geometrical_corrections}, the energy dependence of 2 (out of 25)
detectors of the ROSPHERE is shown as a representative example. These corrections
are implemented in the theoretical angular correlation functions, which are then compared
to the experimental data. When analysing angular correlation measurements for double cascades, the products of the geometrical attenuation coefficients corresponding to the two detectors that register the first and the second $\gamma$ ray must be used, that is $Q_{22}=Q_2(\gamma_1) \times Q_2(\gamma_2)$ and $Q_{44}=Q_4(\gamma_1) \times Q_4(\gamma_2)$ . The subsequent analysis procedure is described thoroughly in the next section.

\section{Analysis method}
\label{sec:analysis}
%
The angular correlation of two $\gamma$ rays involved in a cascade:
\begin{equation}\label{eq:cascade1}
    J_0 \xrightarrow[]{\gamma_0, \delta_0} J_1 \xrightarrow[]{\gamma_1, \delta_1} J_2
\end{equation}
is analysed by performing a least-squares minimization on the experimental data points.
If the spins $J_1, J_2$ and the mixing ratio
$\delta_1$ are known, then the values of $J_0$ and $\delta_0$ can
be determined from the best-fitting theoretical calculation to the experimental data. As discussed in detail in~\cite{ROBINSON1990386}, the correct procedure for the analysis of the angular correlations requires the following function to be formed:
\begin{equation}\label{eq:S_squared}
	S^2 = \sum_i \left[ \frac{W_{i}(\theta^i_{rel})-W_{th}(\delta,\theta^i_{rel}) }{\sigma_{W_i}} \right]^2,
\end{equation}
where $W_i(\theta_{rel}^i)$ are the experimental data points for each relative angle $\theta_{rel}$,
$\sigma_{W_i}$ are the uncertainties of the data and $W_{th}$ are the values of the
theoretical angular correlation function at the same relative angles and corrected for the finite size of the detectors:
\begin{equation}\label{eq: theoretical ang. cor. function}
    W_{th}=A_0 \left[1+ Q_{22}a^{th}_2 P_2(
    \cos \theta_{rel})+Q_{44}a^{th}_4 P_4(\cos \theta_{rel})  \right].
\end{equation}
Here, $A_0$ is a normalization factor which is determined by the experimental data, after a least-squares fit with the same angular correlation function. The factors $Q_{22}$ and $Q_{44}$ are the solid-angle correction factors and they account for the correction due to the finite size of the detectors. The coefficients $a^{th}_2$, $a^{th}_4$ are the theoretically calculated coefficients which
correspond to the values of the spins of the de-excited levels and the
mixing ratios of the corresponding transitions. They are given by the following relations, for $\lambda=2,4$~\cite{Hamilton1975_electromagnetic}:
\begin{equation}\label{eq: a2a4coefficients}
a_\lambda^{th} = B_\lambda(L,L',J_2,J_3,\delta_0) A_\lambda(L,L',J_2,J_1,\delta_1), 
\end{equation}
where $B_\lambda(L,L',J_2,J_3,\delta_0)$ is the orientation parameter of the first transition, $A_\lambda(L,L',J_2,J_1,\delta_1)$ is the angular distribution coefficient referring to the second transition. Both are functions of the mixing ratios $\delta_{0}, \delta_1$ of the two lowest multipolarities $L,L'$ of each $\gamma$ transition. The explicit expressions for the orientation parameters and angular correlation coefficients can be found in~\cite{Hamilton1975_electromagnetic, STUCHBERY200369}, and are given also below for completeness:
\begin{align}\label{eq: orientation parameters}
	B_\lambda(L,L',J_1,J_0,\delta_0) &= \dfrac{1}{1+\delta_0^2} [ F_\lambda(L,L,J_1,J_0) \nonumber \\
&+  (-1)^{L+L'} 2 \delta_0 F_\lambda(L,L',J_1,J_0)  \nonumber \\ 
	&+  \delta_0^2 F_\lambda(L',L',J_1,J_0) ]
	\end{align}
and 
	\begin{align}\label{eq: angular distribution coefficients}
	A_\lambda(L,L',J_1,J_2,\delta_1) &= \dfrac{1}{1+\delta^2_1} [ F_\lambda(L,L,J_1,J_2) \nonumber \\
	&+  2 \delta_1 F_\lambda(L,L',J_1,J_2) \nonumber \\
	&+ \delta^2_1 F_\lambda(L',L',J_1,J_2) ],
	\end{align} 
where the $F_\lambda$ coefficients can be theoretically calculated from explicit expressions found in~\cite{Hamilton1975_electromagnetic, KRANE_STEFFEN_1973_351,Chalil2022} and tabulated in~\cite{Ferentz_F_coeff}. If the second transition is a pure transition of one multipolarity $L$ then $\delta_1=0$ and Eq.~\ref{eq: angular distribution coefficients} is simplified:
\begin{align}\label{eq: angular distribution coefficients simplified}
	A_\lambda(L,J_1,J_2) =  F_\lambda(L,J_1,J_2).
	\end{align} 

The function of Eq.~\ref{eq:S_squared}
is then calculated  for every possible value of the mixing ratio $\delta_0$ of the two lowest
multipolarities $L,L'$. The mixing ratio $\delta_0$ for the transition $\gamma_0$ can then be obtained by varying the $S^2$ function over the mixing ratio $\delta_0$ for all possible initial spin values $J_0$. For each value of the initial spin $J_0$, a curve $S^2$ vs.  $\tan^{-1}(\delta)$ is constructed.
The minimum value of the $S^2$ function with respect to all possible spin values of the initial spin $J_0$ and the mixing ratio $\delta_0$ of the first transition
 will then determine their most probable values. The procedure of assigning the standard error on the mixing ratio is discussed in~\cite{ROBINSON1990386,CLINE1970_291} and is given by the relation:
\begin{equation}\label{eq:internal_error}
S_{lim}^2 = S^2_{min} +1,
\end{equation}
which is used to determine the values of the mixing ratio located at the intersections between the $S^2$ curve and $S_{lim}^2$.
\begin{figure*}[t]
	\centering

	\begin{subfigure}{0.5\textwidth}
		\includegraphics[width=\textwidth]{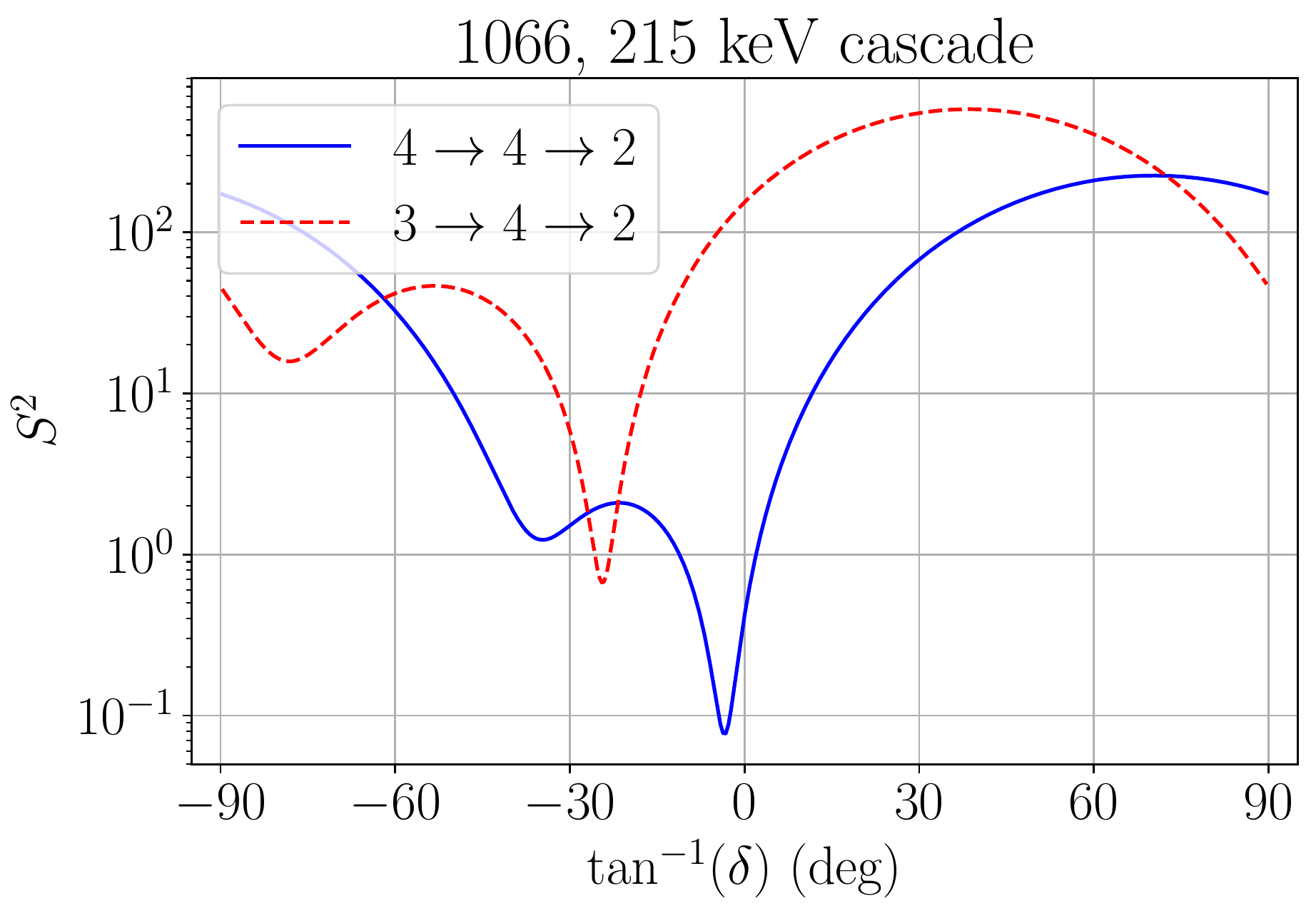}
		\centering
		\caption{}
	\end{subfigure}%
	\begin{subfigure}{0.5\textwidth}
		\includegraphics[width=\textwidth]{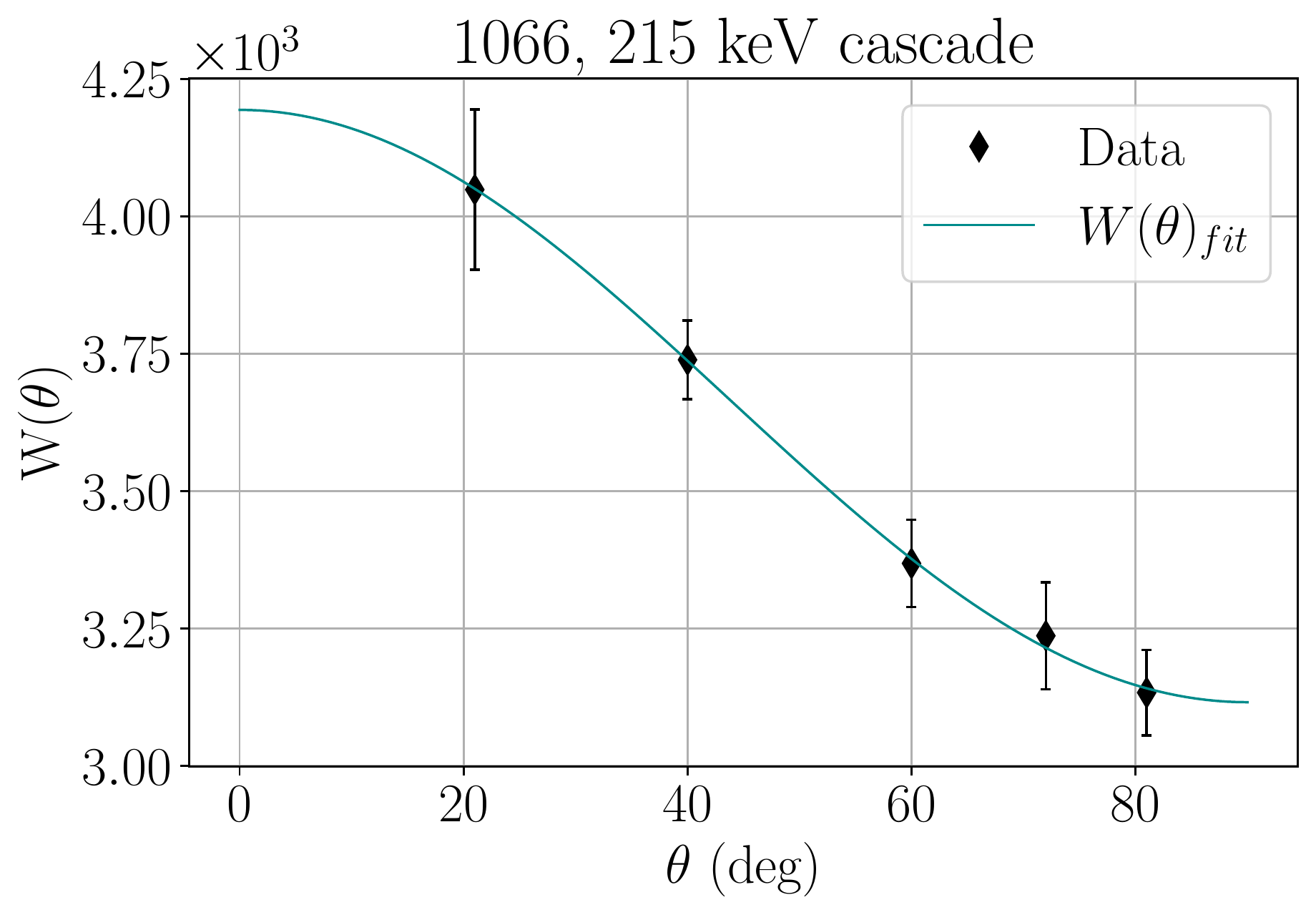}
		\centering
		\caption{}
	\end{subfigure}%
    \caption{The $S^2$ function with respect to the arctangent of the mixing ratio $\delta$
    is shown in (a).
    In (b), the experimental angular correlation function and the best-fit result, using Eq.~\ref{eq: theoretical ang. cor. function}, is shown for the cascade $1374 \rightarrow 309 \rightarrow 93$ keV. The transition energies of the two $\gamma$ rays are shown at the top of the figures.}
	\label{fig:correlation1066_215}
\end{figure*}

It is important to note that the sign of $\delta$, is a matter of convention, as discussed in~\cite{Lange_1982_RevModPhys.54.119}.
Choosing a convention, however, is a non-trivial problem because the sign of the
theoretical $\delta$ depends on the sign conventions employed for defining the
electromagnetic operators and the reduced matrix elements. On the other hand, the
sign of the experimental $\delta$ depends on the sign conventions employed for
defining the axis of alignment with respect to which the $\gamma$--ray angular
distribution is measured, and geometrical factors, such as Clebsch--Gordan and Racah
coefficients that enter the expression employed for the expansion of the angular
distribution probability in terms of various polynomials. The convention used for the measurement in this work is that of Krane and Steffen
in~\cite{KRANE_STEFFEN_1973_351}.

\section{Results and Discussion}
\label{sec:res}

The 1374 keV level, which is the bandhead of the $K^\pi=(4^-)$ two-quasiparticle band (configuration: 9/2[624]-1/2[510])~\cite{NNDC} was assigned a tentative spin of $(4^-)$~\cite{NNDC}, which was the value measured in~\cite{Grigorev1991}. This result is inconsistent with the value of $3^-$ assigned in~\cite{Boneva1990}. The 1374~keV level decays to the $308.6$ keV $4^+_1$ of the rotational ground-state band. The mixing ratio of the transition $1374 \rightarrow 309$ keV has been measured in this work and is compared with the previous value of $\delta(M2/E1) = -0.12(30)$ reported in~\cite{Grigorev1991}.

The results of the angular correlation analysis for the cascade $1374 \rightarrow 309 \rightarrow 93$ keV of \isotope[180][]{Hf} is shown in Fig.~\ref{fig:correlation1066_215}. The two most probable values for the spin of the initial level are $J_0=3,4$ and are tested for their goodness of fit, after correcting for the efficiency of the detectors and for the finite size of the detectors' dimensions. The geometrical attenuation factors are included in the theoretical calculations when the function $S^2$ is formed in Eq.~\ref{eq:S_squared}. The reason for the inclusion of these factors in the theoretical values $W_{th}(\theta)$ is that these corrections cannot be assigned specifically to an experimental point. Since the data are directly compared with theoretical angular distributions, it is more practical to include these factors when calculating the theoretical values for each relative angle $\theta$. The correct implementation of the geometrical attenuation coefficients is important for the final result, as their values can be relatively large depending on the distances and the power of the fold of the coincident data.

The absolute minimum of the $S^2$ vs. $  \tan^{-1} \delta$ curve, as shown in
Fig.~\ref{fig:correlation1066_215}, indicates that the most probable spin value is 4, in agreement with the value measured
in~\cite{Grigorev1991}.  The most probable value for the mixing ratio of quadrupole to dipole has been found in this work equal to $\delta(M2/E1)=-0.06^{+0.10}_{-0.15}$. This value corresponds to the following values for the theoretical angular correlation coefficients: $a_2^{th}=0.213$ and $a_4^{th}=0.001$. 

The mixing ratio measured in the present work is in agreement with the mixing ratio measured in~\cite{Grigorev1991}, found equal to $-0.12(30)$, while significantly improving on the relative uncertainty of the previous measurement. The present measurement cannot fully exclude the possible spin value of $3^-$~\cite{Boneva1990}, as it is below the 95\% confidence interval, which is equal to $S^2_{0.95}=7.8$ for the present case. However, this value seems much less probable with the present data, as the respective minimum of the $3\rightarrow4 \rightarrow 2$ curve is much higher than the minimum of $4 \rightarrow 4 \rightarrow 2$ curve, as illustrated in in Fig~\ref{fig:correlation1066_215}. The 1374 keV level can be assigned a spin and parity of $4^-$, as clearly favored from the present data. In terms of the type of radiation, the low value of the mixing ratio favors that the two lowest multipolarities are of $E1 + M2$ character. The negative sign of the mixing ratio also agrees with the previous measurements~\cite{Grigorev1991,Boneva1990}. As observed in past measurements, negative signs are common for transitions of $E1 + M2$ character~\cite{Uluer_1975}. The compatibility of the mixing ratio with zero though, can support also a pure $E1$ transition.

Overall, the present work provides a new measurement for the mixing ratio of the $1374 \rightarrow 309$ keV transition, whose improved uncertainty compared to previous works can be used for constraining theoretical models, especially for the case of isomeric states such as the 1374 keV state, which is depopulated by the measured transition. The establishment of a spin-parity value of $4^-$ constitutes a firm conclusion after the present results, which should resolve any previous ambiguities in the next ENSDF evaluation for \isotope[180]{Hf}.

\section*{Acknowledgments}

AC and AZ acknowledge support by the Hellenic Foundation for Research and Innovation (HFRI)
and the General Secretariat for Research and Technology (GSRT) under the HFRI PhD Fellowship
grant (GA. No. 74117/2017 and 101742/2019, respectively). Partial support from ENSAR2
(EU/H2020 project number: 654002) is acknowledged. The authors are thankful to the
staff of the 9MV Tandem Laboratory at Horia Hulubei National Institute of Nuclear
Physics and Engineering for both their scientific and technical support during the
experiment.

\bibliography{180Hf.bib}
\end{document}